\documentclass[12pt]{article}
\newcommand{\Comment}[1]{{}}
\usepackage[textwidth = 430 pt, textheight = 630 pt]{geometry}
\usepackage{amssymb,euscript,amsmath,amsfonts}

\usepackage{color}
\definecolor{MyDarkBlue}{rgb}{0.15,0.15,0.45}
 \usepackage[linktocpage=true]{hyperref}  
\hypersetup{
colorlinks=true,
citecolor=MyDarkBlue,
linkcolor=MyDarkBlue,
urlcolor=MyDarkBlue,
pdfauthor={Neil Lambert },
pdftitle={ }, 
pdfsubject={hep-th}
}

\newcommand\ignore[1]{}
\def\one{{\,\hbox{1\kern-.8mm l}}}

\newcommand{\Cset}{{\,\,{{{^{_{\pmb{\mid}}}}\kern-.45em{\mathrm C}}}}}

\newcommand{\nn}{\nonumber}

\newcommand{\be}{\begin{equation}}
\newcommand{\ee}{\end{equation}}
\newcommand{\bea}{\begin{eqnarray}}
\newcommand{\eea}{\end{eqnarray}}

\numberwithin{equation}{section}
\begin{document}

\renewcommand{\thefootnote}{\fnsymbol{footnote}}

   \vspace{1.8truecm}

\vskip 2cm

 \centerline{\LARGE \bf {\sc Non-Lorentzian Field Theories}} 
 \vskip12pt
 \centerline{\LARGE \bf {\sc with Maximal Supersymmetry} } \vskip12pt
 \centerline{\LARGE \bf {\sc and Moduli Space Dynamics} }
 \vspace{2truecm} \thispagestyle{empty} 
 \centerline{
    {\large {\bf {\sc Neil Lambert}}}\footnote{ {\tt neil.lambert@kcl.ac.uk}}  
      {and} {\large {\bf{\sc Miles Owen}}}\footnote{ 
                                  {\tt miles.owen@kcl.ac.uk} }
                                                           }

\vspace{1cm}
\centerline{ {\it Department of Mathematics}}
\centerline{{\it King's College London}}
\centerline{{\it The Strand}}
\centerline{{\it London WC2R 2LS}}
\centerline{EU}
 \vspace{1.0truecm}

 
\thispagestyle{empty}

\centerline{\sc Abstract}
\vspace{0.4truecm}
\begin{center}
\begin{minipage}[c]{360pt}{
We present lagrangian  gauge theories in 2+1 and 4+1 dimensions with 16 supersymmetries which are invariant under rotations and translations but not boosts. The on-shell conditions  reduce the dynamics to motion on a moduli space of BPS states graded by a topologically conserved quantity. On each component of the moduli space only half the supersymmetry is realised. We argue that these theories describe M2-branes and M5-branes which have been infinitely boosted so that their worldvolume `time' has become null.
}

\end{minipage}
\end{center}

\newpage 
 
\renewcommand{\thefootnote}{\arabic{footnote}}
\setcounter{footnote}{0} 

\section{Introduction}

M-theory and its branes are not yet particularly well understood. And their study promises to teach us a great deal about supersymmetric quantum field theory more generally. Of particular interest is the elusive six-dimensional $(2,0)$-Theory of $N$ M5-branes \cite{Witten:1995zh,Strominger:1995ac}. Not only does the formulation of this theory provide an important challenge to our understanding of quantum field theory it's existence also unifies and explains many non-perturbative aspects of lower-dimensional quantum field theories. 

For a variety of reasons it is believed that there is no six-dimensional diffeomorphism invariant lagrangian formulation of the $(2,0)$-Theory ({\it e.g.} see \cite{Witten:1996hc}). However there are a myriad of lagrangians that are associated to lower-dimensional compactifications that can capture some, or even all, of the $(2,0)$-Theory dynamics. In particular when reduced on a circle of radius $R$ the $(2,0)$-Theory becomes   five-dimensional maximally supersymmetric Yang-Mills theory with gauge group $U(N)$ and  coupling $g^2 = 4\pi^2 R$. Alternatively one can think of the $(2,0)$-Theory as providing a   strong coupling, UV completion of the perturbatively non-renormalizable five-dimensional Yang-Mills theory  \cite{Seiberg:1997ax}.

In  \cite{Lambert:2010wm,Lambert:2016xbs} a non-abelian system of equations was formulated which provide a representation of the six-dimensional $(2,0)$ superalgebra. The system involves a set of dynamical equations as well as some constraint equations. Solving the constraints in different ways leads to maximally supersymmetric Chern-Simons theory in 2+1 dimensions or maximally supersymmetric Yang-Mills in 4+1 dimensions, corresponding to M2-branes and M5-branes on $S^1$ respectively.

Owing to the manifest Lorentz symmetry of the system there is also the possibility to construct limits of M2-branes and M5-branes which have been infinitely boosted along some direction (off the brane for M2's but on the brane for M5's). These equations were analysed in \cite{Kucharski:2017jwv,Lambert:2011gb} for M2-branes and M5-branes respectively  and shown to reduce to motion on a moduli space of solitons. In the latter case this reproduces the DLCQ description of the $(2,0)$-Theory as motion on the moduli space of self-dual gauge fields \cite{Aharony:1997th,Aharony:1997an}. We refer to these infinitely boosted branes as null-branes due to the fact that their worldvolume `time' coordinate is  null direction in spacetime. 
 
The main point of this paper is to show that  for these null cases we can construct lagrangians for the dynamics. The result are novel field theories in 2+1 and 4+1 dimensions with 16 supersymmetries, translations and spatial rotations  but which are not invariant under boosts.\footnote{ More preceisely there is no manifest boost symmetry. It is conceivable that boosts can still be identified in non-linear and potentially non-local manner.}  Furthermore the field content includes non-dynamical Lagrange multiplier fields which restrict the dynamics to motion on a moduli space of solitons. These appear  to be a new type of maximally supersymmetric lagrangian and it would be interesting to seek other examples.

In the rest of this paper we will study the M2-brane example in section two. This is a field theory in 2+1 dimensions with maximal supersymmetry, an $SO(2)$ rotational symmetry and an $SO(2)\times SO(6)$ R-symmetry. In section three we construct the M5-brane example which is a field theory in 4+1 dimensions with maximal  supersymmetry, an $SO(4)$ rotational symmetry and $SO(5)$ R-symmetry. We also briefly explore the dimensional reduction of these theories.  In section four we comment on how the 8 supersymmetries that arise in the moduli space dynamics are enhanced to 16 supersymmetries in the field theory. In the final section we give our conclusions and comments. We also provide an appendix with some useful Fierz identities.
  
 \section{The Null M2}

In \cite{Kucharski:2017jwv} a novel system of equations was derived from the system of \cite{Lambert:2016xbs}. In particular the dynamical  fields  take values in a three-algebra with invariant inner-product $\langle\ \cdot\ ,\ \cdot\ \rangle $ and totally anti-symmetric product:
\begin{equation}
[\ \cdot\ ,\ \cdot\ ,\ \cdot\ ]:{\cal V}\otimes {\cal V}\otimes {\cal V} \to {\cal V}\ ,
\end{equation}
that satisfies the fundamental identity
\begin{equation}
[U,V,[X,Y,Z]] = [[U,V,X],Y,Z] +[[X,[U,V,Y],Z] +[X,Y,[U,V,Z]] \ .
\end{equation}
 In addition the three-algebra generates a Lie-algebra ${\cal G}$  by the three-algebra analogue of the adjoint map: $X\to\varphi_{U,V}(X) = [U,V,X]$ 
defined by any pair $U,V\in{\cal V}$. This also induces an invariant  Lie-algebra inner-product $(\ \cdot\ ,\ \cdot\ )$ on  ${\cal G}$ that satisfies
 \begin{equation}
 (T,\varphi_{U,V}  ) = \langle T(U),V\rangle  \ ,
 \end{equation}
for any $T \in {\cal G}$ and pair $U,V\in{\cal V}$. More concretely there is a unique finite-dimensional three-algebra  with positive definite inner-product $\langle\ \cdot\ ,\ \cdot\ \rangle $ \cite{Gauntlett:2008uf,Papadopoulos:2008sk}. In particular ${\cal V}= {\mathbb R}^4$ with orthonormal basis $T^A$, $A=1,2,3,4$ and
\begin{equation}
[T^A,T^B,T^C] = \frac{4\pi}{k}\varepsilon^{ABCD}T^D\ .
\end{equation}
In which case one finds ${\cal G}= su(2)\oplus su(2)$ and  $(\ \cdot\ ,\ \cdot \ )$ acts as $(k/4\pi) {\rm tr}, -(k/4\pi){\rm tr}$ on the two $su(2)$ factors respectively. 

Some trial and error shows that the equations of motion of \cite{Kucharski:2017jwv}
arise from the action\footnote{Note that compared to \cite{Kucharski:2017jwv}  we have rescaled
$X^I\to l^{-3/2}X^I , Z \to 2l^{3/2}Z, H\to \frac12  l^{-3/2}H  , \Psi_{\pm} \to l^{-3/2}\Psi_{\pm}$ so that the fields have canonical scaling dimensions.}
\begin{equation}\label{M2action}
S_{M2} = S_{scalar}+S_{CS}+S_{fermion}\ ,
\end{equation}
where
 \begin{align}
S_{scalar} &= \int d^2x dt\left[ \langle D_tZ,D_t\bar Z\rangle 
 -   \langle DX^I,\bar DX^I\rangle+  \langle D\bar Z,  \bar H \rangle +   \langle \bar DZ,H\rangle\right.\nonumber\\
&\qquad \qquad \left. -i\langle D_tX^I,[Z,\bar Z, X^I] \rangle   -\frac{1}{  2}\langle [X^I,X^J,Z][X^I,X^J,\bar Z]\rangle  \right] 
\nonumber\\
S_{CS} & = i\int d^2x dt\left[ \frac12(A_t, F_{z\bar z}) + \frac12(A_z, F_{\bar z t}) +\frac12(A_{\bar z}, F_{t z}) +\frac{1}{2}(A_t,[A_z,A_{\bar z}])\right]\nonumber\\
S_{fermion} &= \int d^2x dt\left[ \frac{i }{2\sqrt{2}}\langle \Psi_+^T,D_t\Psi_+\rangle+i  \langle \Psi_+^T,\hat\Gamma_z\bar D\Psi_- + \hat\Gamma_{\bar z}D\Psi_-\rangle  \right.\nonumber\\
&\qquad\left.-\frac{ 1}{2\sqrt{2}}\langle\Psi_+^T ,  \hat{\Gamma}_{Z \bar{Z}} \hat{\Gamma}^{IJ} \left[ X^I, X^J, \Psi_+ \right]\rangle+\frac{1}{ \sqrt{2}} \langle\Psi_-^T,\left[ Z, \bar{Z}, \Psi_- \right]\rangle\right.\nonumber\\
&\qquad \left. + i\langle\Psi_+^T ,  \hat{\Gamma}^I \hat{\Gamma}_Z \left[ Z, X^I, \Psi_- \right] \rangle+ i\langle\Psi_+^T ,  \hat{\Gamma}^I \hat{\Gamma}_{\bar{Z}} \left[ \bar{Z}, X^I, \Psi_- \right] \rangle\right]\ .
\end{align}
Here the dynamical fields 
consist of six scalars $X^I$, $I=5,6,7,8,9,10$, a complex scalar $Z = X^4+iX^3$ and fermions $\Psi_{\pm}$ satisfying
\begin{align}
\hat\Gamma_{t12}\Psi_\pm =-\Psi_\pm\qquad \hat\Gamma_{t34}\Psi_\pm =\pm\Psi_\pm \ .
\end{align}
where $\hat \Gamma_t,\hat\Gamma_1,...,\hat\Gamma_{10}$ form a real basis of the $Spin(1,10)$ Clifford algebra\footnote{The hat arises as this basis is adapted from M5-branes to M2-branes in the construction of \cite{ Lambert:2016xbs,Kucharski:2017jwv} but this distinction is not necessary here and can be dropped. }. In addition we have introduced the complex coordinate $z = x^1+ix^2$ and 
\begin{equation}
\hat\Gamma_Z = \frac{1}{2}(\hat\Gamma_3 - i\hat\Gamma_4)\qquad \hat\Gamma_z = \frac{1}{2}(\hat\Gamma_1 - i\hat\Gamma_2)\ .
\end{equation}
There is also a non-dynamical complex scalar $H$ which also takes values in the three-algebra ${\cal V}$ and a one-form gauge field $(A_t,A_z,A_{\bar z})$ taking values in the associated Lie-algebra ${\cal G}$. 

The action has an $SO(2)$ rotational symmetry along with an  $SO(2)\times SO(6)$ R-symmetry. In addition   
one can explicitly check that it is also invariant under the  sixteen supersymmetries derived in \cite{Kucharski:2017jwv}:
\begin{align}
\delta X^I =& i \epsilon^T_+ \hat{\Gamma}^I \Psi_- + i \epsilon^T_- \hat{\Gamma}^I \Psi_+\nn\\
\delta Z =&  \sqrt{2}  \epsilon^T_+ \hat{\Gamma}_{\bar{Z}} \Psi_{+}
\nn\\
\delta \bar{Z} =& -   \sqrt{2}  \epsilon^T_+ \hat{\Gamma}_{Z} \Psi_{+}
\nn\\
\delta A_{z}(\ \cdot \ ) =& \sqrt{2}\epsilon^T_+\hat\Gamma^I\hat \Gamma_z[X^I,\Psi_+,\ \cdot\ ] +  2i \epsilon^T_- \hat{\Gamma}_z \hat{\Gamma}_{\bar{Z}} \left[ \bar{Z}, \Psi_+,\ \cdot\  \right] - 2i \epsilon^T_+ \hat{\Gamma}_z \hat{\Gamma}_{Z} \left[ Z, \Psi_-,\ \cdot\   \right]
\nonumber\\
\delta A_{\bar{z}} =& -\sqrt{2}   \epsilon_+^T \hat{\Gamma}^I \hat{\Gamma}_{\bar{z}} [X^I, \Psi_+, \cdot] +   2i \epsilon^T_- \hat{\Gamma}_{\bar{z}} \hat{\Gamma}_{Z} \left[ Z, \Psi_+,\ \cdot \ \right] -2 i \epsilon^T_+ \hat{\Gamma}_{\bar{z}} \hat{\Gamma}_{\bar{Z}} \left[ \bar{Z}, \Psi_-,\ \cdot\  \right]
\nn \\
\delta A_t(\ \cdot \ ) =&  2\sqrt{2} i \epsilon^T_- \hat{\Gamma}_Z \left[ Z, \Psi_-,\ \cdot \ \right] + 2\sqrt{2} i \epsilon^T_- \hat{\Gamma}_{\bar{Z}} \left[ \bar{Z}, \Psi_-, \ \cdot\ \right] \nonumber \\  & + 2  \epsilon^T_- \hat{\Gamma}_{Z \bar{Z}} \hat{\Gamma}^I \left[X^I, \Psi_+,\ \cdot\  \right] - 2  \epsilon^T_+ \hat{\Gamma}_{Z \bar{Z}} \hat{\Gamma}^I \left[X^I, \Psi_-,\ \cdot\  \right]\nn\\
\delta \Psi_+  
=&   2i{\sqrt{2} } \hat{\Gamma}^I \left[ Z, \bar{Z}, X^I  \right] \epsilon_- - 2i \left( \hat{\Gamma}_Z D_t Z -  \hat{\Gamma}_{\bar{Z}} D_t \bar{Z}  \right) \epsilon_+ \nonumber \\ & -  \left( \hat{\Gamma}_Z \hat{\Gamma}^{IJ} \left[ Z, X^I, X^J  \right] + \hat{\Gamma}_{\bar{Z}} \hat{\Gamma}^{IJ} \left[ \bar{Z}, X^I, X^J  \right]  \right)\epsilon_+ \nonumber \\ &\  + 2 \left( \hat{\Gamma}_{\bar{z}} \hat{\Gamma}^I D X^I + \hat{\Gamma}_{{z}} \hat{\Gamma}^I \bar{D} X^I  \right) \epsilon_+ \nonumber \\ & +  2{\sqrt{2}i}  \left(  \hat{\Gamma}_{\bar{z}} \hat{\Gamma}_Z D Z - \hat{\Gamma}_z \hat{\Gamma}_{\bar{Z}}\bar D\bar{Z}\right) \epsilon_-\nn\\
\delta \Psi_-  =& -\sqrt{2} \hat{\Gamma}^I D_t X^I \epsilon_+ - \frac{  \sqrt{2}i  }{3}\hat{\Gamma}_{Z \bar{Z}} \hat{\Gamma}^{IJK} \left[ X^I, X^J, X^K  \right] \epsilon_+   \nonumber \\ & +    \left( \hat{\Gamma}_Z \hat{\Gamma}^{IJ} \left[ Z, X^I, X^J  \right]  + \hat{\Gamma}_{\bar{Z}} \hat{\Gamma}^{IJ} \left[ \bar{Z}, X^I, X^J  \right]  \right) \epsilon_- \nonumber \\  
&+ 2 \left( \hat{\Gamma}_{\bar{z}} \hat{\Gamma}^I D X^I + \hat{\Gamma}_{{z}} \hat{\Gamma}^I \bar{D} X^I  \right) \epsilon_- \nonumber\\
&-2i \left( \hat{\Gamma}_Z D_t Z - \hat{\Gamma}_{\bar{Z}} D_t \bar{Z}  \right) \epsilon_- +  \sqrt{2}i  \left( \hat{\Gamma}_{\bar{z}}\hat{\Gamma}_{\bar{Z}} H - \hat{\Gamma}_z  \hat{\Gamma}_Z \bar{H} \right)  \epsilon_+\ .\nonumber\\
\delta H =
&   2\sqrt{2}  \epsilon^T_-  \hat{\Gamma}_Z    D\Psi_- + 2\epsilon^T_+ \hat{\Gamma}_z \hat{\Gamma}_Z D_t \Psi_-  \nn\\
& +  i \epsilon^T_+ \hat{\Gamma}_z \hat{\Gamma}_Z  \hat{\Gamma}^{IJ} \left[ X^I, X^J, \Psi_-  \right] -2\sqrt{2} \epsilon_-^T \hat{\Gamma}_z \hat{\Gamma}^I[\bar Z,X^I,\Psi_-]\ ,
\end{align}
where 
\begin{align}
\hat\Gamma_{t12}\epsilon_\pm =\epsilon_\pm\qquad \hat\Gamma_{t34}\epsilon_\pm =\pm\epsilon_\pm \ .
\end{align}
While examining the cubic fermion terms that arise in $\delta S$ it is helpful to observe that they take the same form as the cubic fermion terms that arise in the case of the maximally supersymmetric Lorentzian M2-brane theory  (see the appendix).

The action (\ref{M2action}) has some non-standard features. Firstly although the scalars $Z$ have canonical kinetic terms they do not have gradient terms. The scalars $X^I$ have the opposite: no kinetic terms but canonical gradient terms. Furthermore there is a term which is linear in the $X^I$ time-derivative. 

We see that the field $H$ imposes a holomorphic constraint
\begin{equation}
\bar D Z=0\ .
\end{equation}
We also have the Gauss law constraint arising from the $A_t$ equation of motion:
\begin{equation}
F_{z \bar{z}} (\cdot)  = -i \left( \left[Z, D_t \bar{Z},\ \cdot\  \right] + \left[ \bar{Z}, D_t Z, \ \cdot \ \right] \right)   -     \left[ X^I, \left[Z, \bar{Z}, X^I  \right],  \ \cdot\ \right] - \frac{1}{2\sqrt{2}} \left[ \Psi_+^T, \Psi_+,\ \cdot\  \right] \  .
\end{equation}
For static bosonic configurations these constraints reduce to a 3-algebra form of the Hitchin System:
\begin{align}
\bar D Z&=0\nonumber\\ 
F_{z \bar{z}} (\cdot)  &=  -     \left[ X^I, \left[Z, \bar{Z}, X^I  \right],  \ \cdot\ \right]   \  .
\end{align}
These arise as BPS solutions to   M2-brane \cite{Kim:2009ny}. It was shown in \cite{Kucharski:2017jwv} that allowing for time evolution the dynamical evolution is still restricted to the Hitchin moduli space (at least for a class of configurations). 

Finally we note that in 
 \cite{Kucharski:2017jwv} this system was identified as describing intersecting M2-branes along the $x^1,x^2$ and $x^3,x^4$ directions, in the limit of an infinite boost along $x^5$. The $SO(2)\times SO(6)$ R-symmetry then arises from rotations in the two-dimensional $(x^3,x^4)$-plane of the  M2-brane and in the six-dimensional plane orthogonal to both M2-branes respectively.

\section{Null M5-branes}

We now turn our attention to a similar construction that represents M5-branes. 
Although the system is also derived from the three-algebra construction of \cite{Lambert:2010wm} it turns out that the resulting dynamical equations can be extended to any gauge group (for example by considering a non-positive definite three-algebra and decoupling the negative definite modes).  In particular the field content consists of five scalars $X^I$ (where now $I=6,7,8,9,10$), a gauge field one-form $(A_t,A_i)$, $i=1,2,3,4$   and fermions $\Psi$ all taking values in some Lie-algebra. There is also an anti-self-dual tensor $G_{ij}$.  We consider the action
\begin{eqnarray}\label{SM5}
S_{M5} &=& \frac{1}{ g^2} \text{tr} \int \; d^4x dt \left( \frac{1}{2 } F_{ti}F_{ti} - \frac{1}{2}D_i X^I D_i X^I + \frac{1}{2 }F_{ij}G_{ij} \right. \nonumber \\ &\phantom{=}& \phantom{\text{tr}\int \; d^4x dt} \left. + \frac{i }{2} \bar{\Psi}\Gamma_- D_t \Psi  + \frac{i }{2} \bar{\Psi} \Gamma_i D_i \Psi -\frac{1}{2} \bar{\Psi} [X^I, \Gamma_- \Gamma^I\Psi] \right)\ , 
\end{eqnarray} 
where    $\bar \Psi= \Psi^T\Gamma_t$.
Here the fermions satisfy $\Gamma_{t12345}\Psi=-\Psi$ and we define
\begin{equation}
\Gamma_\pm  = \frac{1}{\sqrt{2}}(\Gamma_5\pm\Gamma_t)\ .
\end{equation}
Again $\Gamma_t,\Gamma_1,...,\Gamma_{10}$ are a real representation of the $Spin(1,10)$ Clifford algebra. Note that, unlike the gauge field strength $F_{ij}$, $G_{ij}$ does not satisfy a Bianchi identity.

The equations of motion arising from this action agree with those constructed in \cite{Lambert:2010wm,Lambert:2011gb}\footnote{Here we have rescaled  the fields  from those of reference \cite{Lambert:2010wm} to their canonical form and also switched the roles of $t=-x^+$ and $x^-$.}.
In particular we see that $G_{ij}$ acts as a Lagrange multiplier imposing self-duality of the spatial components of the gauge field strength; $F_{ij} = \frac 12\varepsilon_{ijkl}F_{kl}$.
Thus the on-shell condition reduces to motion on the moduli space of self-dual gauge fields. In particular the action reduces to a sigma-model on ADHM moduli space which includes a potential and background gauge field that arise from the vacuum expectation values of $X^I$ and $A_0$ respectively \cite{Lambert:2011gb}. This agrees with the DLCQ prescription for the M5-brane $(2,0)$ SCFT given in \cite{Aharony:1997th,Aharony:1997an}

First we begin with the  supersymmetries  of  \cite{Lambert:2010wm,Lambert:2011gb}:
\begin{eqnarray}\label{onshellsusy}
\delta X^I &=& i \bar{\epsilon} \Gamma^I \Psi \nonumber \\ \delta A_i &=&  i \bar{\epsilon}\Gamma_i \Gamma_- \Psi \nonumber \\ \delta A_t &=&   i \bar{\epsilon} \Gamma_{+-} \Psi \nonumber \\ 
\delta \Psi &=&  \Gamma_- \Gamma^I  D_t X^I \epsilon + \Gamma_i \Gamma^I  D_i X^I \epsilon +  \Gamma_i \Gamma_{+- } F_{ti} \epsilon  - \frac{1}{4 } \Gamma_+ \Gamma_{ij}  F_{ij} \epsilon \nonumber \\ &\phantom{=}&  - \frac{1}{4 } \Gamma_- \Gamma_{ij}  G_{ij}\epsilon - \frac{i }{2} \Gamma_- \Gamma^{IJ}  [X^I, X^J] \epsilon\nonumber\\
\delta G_{ij} & =&   i\bar\epsilon  \Gamma_{ij}  D_t\Psi +  2i\bar\epsilon\Gamma_+\Gamma_{[i}D_{j]}\Psi  -  \bar\epsilon \Gamma_{ij}\Gamma_{+-}\Gamma^I[X^I,\Psi]\ ,
\end{eqnarray}
where $\Gamma_{t12345}\epsilon=\epsilon$. 
These transformations close on-shell and one can check that the resulting equations of motion are invariant.

 However   to construct a supersymmetry of the action we need to find an  expression for $\delta G_{ij}$ that is anti-self-dual off-shell. Thus the transformations (\ref{onshellsusy}) require some modification. First we observe that we are free to modify  $\delta G_{ij}$ by 
\begin{equation}
\delta G_{ij}\to \delta G_{ij} + i\bar\epsilon \Xi\Gamma_{ijk}D_k\Psi\ ,
\end{equation}
for any choice of $\Xi$, because the Bianchi identity of $F_{ij}$ ensures that the change in $\delta S$ is a boundary term. In particular taking $\Xi = \frac{3}{2}\Gamma_+$ we find that 
\begin{align}
\delta G_{ij} +\star\delta G_{ij}&=   2\bar\epsilon \Gamma_+\Gamma_{ij} E_\Psi\ ,
\end{align}
where 
\begin{equation}
E_\Psi=
i\Gamma_-D_t\Psi + i\Gamma_kD_k\Psi - \Gamma_-\Gamma^I[X^I,\Psi]\ .
\end{equation}
is the fermion equation of motion. 
We can correct this   by making the following shift in the supersymmetry transformations:
\begin{align}
\delta G_{ij} &\to \delta G_{ij}- \bar\epsilon \Gamma_+\Gamma_{ij} E_\Psi
\nonumber\\
\delta \bar \Psi &\to \delta\bar\Psi + \frac{1}{2 }\bar\epsilon \Gamma_+\Gamma_{ij} F_{ij}\ ,
\end{align}
so that the action remains invariant but now $\delta G_{ij}$ is anti-self dual off-shell. One can now see that the action is indeed invariant under 
the following   supersymmetry transformations
 \begin{eqnarray}
\delta X^I &=& i \bar{\epsilon} \Gamma^I \Psi \nonumber \\ \delta A_i &=&  i \bar{\epsilon}\Gamma_i \Gamma_- \Psi \nonumber \\ \delta A_t &=&   i \bar{\epsilon} \Gamma_{+-} \Psi \nonumber \\ 
\delta \Psi &=&   \Gamma_- \Gamma^I  D_t X^I \epsilon + \Gamma_i \Gamma^I  D_i X^I \epsilon +  \Gamma_i \Gamma_{+- } F_{ti} \epsilon  +\frac{1}{4 } \Gamma_+ \Gamma_{ij}  F_{ij} \epsilon \nonumber \\ &\phantom{=}&  - \frac{1}{4 } \Gamma_- \Gamma_{ij}  G_{ij}\epsilon - \frac{i }{2} \Gamma_- \Gamma^{IJ}  [X^I, X^J] \epsilon\nonumber\\
\delta G_{ij} & =& -\frac{i}{2}\bar\epsilon\Gamma_k\Gamma_{ij}\Gamma_+D_k\Psi -\frac{i }{2}\bar\epsilon\Gamma_{-} \Gamma_{ij} \Gamma_+D_t\Psi -\frac12 \bar\epsilon\Gamma_- \Gamma_{ij}\Gamma_+\Gamma^I[X^I,\Psi]\ .
\end{eqnarray}
 When checking the vanishing of the cubic fermion terms in $\delta S$ is it helpful to observe that they have a similar structure to those that arise in maximally supersymmetric five-dimensional Yang-Mills (see the appendix). 
 
Lastly we note that we are free  to add an $F_{ij}F_{ij}$ term into the action:
\begin{equation}
S  \rightarrow S  - \frac{\xi}{4g^2}\int d^4x dt \  F_{ij}F_{ij}\ ,
\end{equation}
for any choice of  $\xi$.
 This will not change the equations of motion since $D^iF_{ij}=0$ as a result of the self-dual condition imposed by $G_{ij}$ along with the Bianchi identity of $F_{ij}$.  Furthermore, to preserve supersymmetry, we simply shift the  variation $\delta G_{ij}$ to
\begin{equation}
\delta G_{ij} \rightarrow \delta G_{ij} + 2i\xi  \bar{\epsilon} \Gamma_- \Gamma_{[i} D_{j]} \Psi\ ,
\end{equation}
so as to ensure $\delta S=0$. However in the rest of this paper we will set $\xi=0$ since on-shell  $\xi\ne 0 $ leads to an infinite contribution  to the action arising from the integral over time of the   constant instanton number. 

 In \cite{Lambert:2010wm,Lambert:2011gb} the equations of motion arising from (\ref{SM5}) were interpreted as the limit of an infinite boost of M5-branes along a worldvoume direction $x^5$ with a fixed value for the null momentum $P_-$. In particular preserving $P_-$ breaks the $SO(1,5)$ Lorentz symmetry of the M5-brane worldvolume to $SO(4)$ and leaves the $SO(5)$ R-symmetry and sixteen supersymmetries intact. This agrees with the $SO(4)\times SO(5)$ symmetry and maximal supersymmetry of the action (\ref{SM5}).

\subsection{Dimensional Reduction} 

The action (\ref{SM5}) provides a non-Lorentz invariant field theory in $4+1$ dimensions which is invariant under sixteen supersymmetries, an $ISO(4)$  Euclidean group and an $SO(5)$ R-symmetry. It's on-shell conditions reduce to motion on the moduli space of self-dual gauge fields on ${\mathbb R}^4$ with $t$ playing the role of time. 

Clearly we can dimensionally reduce this action to obtain similar ones in $d+1$ dimensions with $d<4$. Following the usual rules of dimensional reduction over $4-d$ dimensions the bosonic field content is now
\begin{align}
(A_t,A_i)\qquad (X^a = A_{d+1},..,A_4)\qquad  (X^I) \qquad (G_{ij}, G_{ia}, G_{ab})\ ,
\end{align}
where  now the $i$ index has been reduced to  $i=1,..,d$ with  $a = d+1,..,4$ and as before we have $I=6,7,8,9,10$. Note also that anti-self-duality implies that the various components  $ (G_{ij}, G_{ia}, G_{ab})$ are not independent. In all these cases the on-shell conditions imply that the dynamics corresponds to motion on the moduli space of  self-dual connections reduced to ${\mathbb R}^{4-d}$.  

One readily sees from (\ref{SM5}) that scalars $X^a$ will have kinetic terms but $X^I$ will not. Furthermore there will be a potential of the form
\begin{equation}
V\sim -{\rm tr}([X^a,X^I][X^a,X^I])\ ,
\end{equation}
but no  potential terms with only $X^I$ or $X^a$.  Thus, unlike the  dimensional reduction of Lorentzian maximally supersymmetric Yang-Mills theories, the R-Symmetry is not enhanced to $SO(9-d)$. Rather, upon reduction to $d+1$ dimensions, we obtain a maximally super-symmetric field theory with 
$ISO(d)$ Euclidean symmetry  and a $SO(4-d)\times SO(5)$ R-symmetry.  

For the sake of completeness let us list the dimensional reductions. 

\noindent {\bf Reduction to 3+1 Dimensions}

 Reduction to 3+1 dimensions we have ($i,j=1,2,3$)
\begin{equation}
(A_t, A_i) \qquad (X^4 \equiv   A_4)\qquad ( X^I) \qquad (G_{ij},  G_{i 4})\ .
\end{equation}
However, we are taking that $G$ is anti-self-dual so we have the relationship
\begin{equation}
G_{ij} = -  \varepsilon_{ijk} G_{k4}  \ .
\end{equation}
Thus the action becomes
\begin{align}
S_{3+1 } &= \frac{1}{ g^2}\int  d^3x dt\left[  \frac{1}{2 } F_{t i  } F_{t i } + \frac{1}{2}D_t X^4 D_t X^4 + \frac{1}{2 } G_{ij} \left( F_{ij}  - \varepsilon_{ijk} D_kX^4 \right) \right. \nonumber \\ & \qquad \qquad \qquad  -\frac{1}{2} D_i X^I D_i X^I + \frac{1}{2} [X^4, X^I][X^4, X^I] \nonumber \\ & \qquad \qquad \qquad +\left. \frac{i}{2} \bar{\Psi} \Gamma_- D_t \Psi + \frac{i }{2} \bar{\Psi} \Gamma_i D_i \Psi + \frac{1}{2} \bar{\Psi} \Gamma_4 [X^4, \Psi] - \frac{1}{2}\bar{\Psi} [X^I, \Gamma_- \Gamma^I \Psi] \right]\ .
\end{align}

\noindent {\bf Reduction to 2+1}

Next let us look at the   reduction to $2+1$ dimensions and compare the result with the action in section two.    The field content is given by ($i=1,2$, $a=3,4$)
\begin{equation}
(A_t, A_i) \qquad (X^a \equiv   A_a)\qquad (X^I ) \qquad (G_{ij}, G_{ab}, G_{ia})\ ,
\end{equation}
but due to anti-self-duality the components $G_{ij}$ and $G_{ab}$ are related as are the various components of $G_{ia}$.   Let us introduce the complex coordinates
\begin{equation}
z = x^1 + ix^2 \qquad Z = X^4 +iX^3\ ,
\end{equation}
and
\begin{equation}
D = \frac12(D_1 - iD_2) \qquad \Gamma_Z = \frac{1}{2}(\Gamma_4 - i\Gamma_3)\ .
\end{equation}
We also re-express the independent components of the Lagrange multiplier field as
\begin{equation}
G=G_{12}=-G_{34}\qquad H = G_{14}-iG_{13}\ .
\end{equation}
We these definitions we can write the reduced action as
\begin{eqnarray}\label{SD2}
S_{2+1}&=&\frac{1}{ g^2}{\rm tr}\int \, d^2xdt \left( \frac{1}{2 }F_{t z  } F_{t \bar z   } + \frac{1}{2} D_t Z D_t \bar{Z}  + \bar H D\bar{Z} + H\bar{D}Z      \right. \nonumber \\ &\phantom{=}& \phantom{\int \, d^2x dt} -  D X^I \bar D X^I + \frac{1}{2}[Z,X^I][\bar{Z}, X^I] -2iG\left(   F_{z\bar z}  - \frac{1}{4}[Z, \bar{Z}] \right) \nonumber \\ &\phantom{=}& \phantom{\int \, d^2x dt} +\frac{i}{2} \bar{\Psi} \Gamma_-D_t\Psi + i\bar{\Psi}(\Gamma_{\bar z} D  \Psi+\Gamma_{ z} \bar D  \Psi) + \frac{1}{2} \bar{\Psi} \Gamma_Z [Z, \Psi] + \frac{1}{2}\bar{\Psi} \Gamma_{\bar{Z}}[\bar{Z}, X^I] \nonumber \\ &\phantom{=}& \phantom{\int \, d^2x dt} \left. - \frac{1}{2} \bar{\Psi}[X^I, \Gamma_-\Gamma^I \Psi] \right)\ .
\end{eqnarray}

 The on-shell conditions now reduce to motion on the moduli space of solutions to the Hitchin System,   this time for any gauge group. However although it has the same number of supersymmetries as the M2-brane case discussed above it only has $SO(2)\times SO(5)$ R-symmetry, not $SO(2)\times SO(6)$. It is natural to postulate that, just as the lorentzian M2-brane theory is the strong coupling limit of (2+1)-dimensional maximally supersymmetric Yang-Mills (which can be viewed as the dimensional reduction of the M5-brane), the null M2-brane theory (\ref{M2action}) is the strong coupling fixed point of the null M5-brane action (\ref{SD2}) in the case of an $SU(2)$ gauge group.

\noindent {\bf Reduction to 1+1 Dimensions}

Next we consider the reduction to 1+1 Dimensions. Here the bosonic fields are ($a=2,3,4$)
\begin{equation}
(A_t, A_1) \qquad (X^a \equiv   A_a)\qquad ( X^I) \qquad (G_{ab}, B_a =  G_{1a })\ .
\end{equation}
However, we are taking that $G$ is anti self-dual so we have the relationship
\begin{equation}
G_{ab} = -  \varepsilon_{abc} B_c  \ .
\end{equation}
The action can be written now as
\begin{align}
S_{1+1 } &= \frac{1}{ g^2}\int   dx dt\left[ \frac{1}{2 }F_{t 1 } F_{t  1  }+    \frac{1}{2}D_t X^a D_t X^a - \frac12 D_1X^ID_1X^I+ \frac{1}{2} [X^a, X^I][X^a, X^I]  \right. \nonumber \\   & \qquad \qquad \qquad \left. -\frac{1}{2 } G_{ab} (\varepsilon_{abc}D_1X^c+i[X^a,X^b]) \right. \nonumber \\ & \qquad \qquad \qquad +\left. \frac{i}{2} \bar{\Psi} \Gamma_- D_t \Psi+\frac{i}{2} \bar{\Psi} \Gamma_1 D_1 \Psi +  \frac{1}{2}\bar{\Psi} \Gamma_a [X^a, \Psi] - \frac{1}{2}\bar{\Psi} [X^I, \Gamma_- \Gamma^I \Psi] \right]\ .
\end{align}
Here we see that the Lagrange multiplier reduces the theory to motion on the moduli space of Nahm's equations. 
 
\noindent {\bf Reduction to 0+1 Dimensions}

Lastly we can consider the case of a reduction to 0+1 dimensions. The bosonic fields are ($a=1,2,3,4$)
\begin{equation}
(A_t ) \qquad (X^a \equiv   A_a) \qquad (X^I) \qquad (G_{ab} )\ ,
\end{equation}
and now $G_{ab}$ is anti-self-dual. 
The action becomes
\begin{align}
S_{0+1 } &= \frac{1}{ g^2}\int    dt\left[   \frac{1}{2}D_t X^a D_t X^a + \frac{1}{2} [X^a, X^I][X^a, X^I] - \frac{i}{2 } G_{ab} [X^a,X^b] \right. \nonumber \\   & \qquad \qquad \qquad +\left. \frac{i}{2} \bar{\Psi} \Gamma_- D_t \Psi + \frac{1}{2}\bar{\Psi} \Gamma_a [X^a, \Psi] - \frac{1}{2}\bar{\Psi} [X^I, \Gamma_- \Gamma^I \Psi] \right]\ .
\end{align} 
This is itself a quantum mechanical model whose on-shell equations of motion reduce it to a sigma model on the moduli space of matrices  that satisfy
\begin{equation}\label{bps}
[X^a,X^b] = \frac 12\varepsilon^{abcd}[X^c,X^d]\ .
\end{equation}
However there are no finite dimensional non-trivial solutions to this system. To see this one observes that the expression
\begin{equation}
V =- {\rm tr}([X^a,X^b][X^a,X^b])
\end{equation}
is positive definite but when evaluated on (\ref{bps}) we find
\begin{equation}
V  
= \frac12\varepsilon^{abcd}{\rm tr}(X^a[X^b,[X^c,X^d]]) \ ,
\end{equation}
which vanishes by the Jacobi identity and hence $[X^a,X^b]=0$.
Nevertheless it might be interesting to explore any applications for this model in terms of the Matrix theory approach to M-theory.

\section{Eight vs Sixteen Supersymmetries}

In the examples above we have constructed field theories in a variety of dimensions which are invariant under sixteen supersymmetries. However the on-shell conditions reduce the dynamics to one-dimensional motion  on a finite-dimensional moduli space of BPS configurations (self-dual gauge fields and their various dimensional reductions). However these moduli spaces are hyper-K\"ahler and as such  the one-dimensional sigma-models describing their dynamics  possess only 8 supersymmetries. What has happened?

To resolve this paradox we observe that the  the sixteen supersymmetries split into $({\cal Q}_+,{\cal Q}_-)$ and their algebra takes the form \cite{Kucharski:2017jwv,Lambert:2011gb}
\begin{align}
\{{\cal Q}_+,{\cal Q}_+\} &\sim {\cal P}_+\nonumber\\
\{{\cal Q}_+,{\cal Q}_-\} &\sim {\cal P}\nonumber\\
\{{\cal Q}_-,{\cal Q}_-\} &\sim {\cal P}_- \ .
\end{align}
Here ${\cal P}_+$ is the  energy arising from the lagrangians above, ${\cal P}$ denote the spatial momenta and ${\cal P}_-$ is a topological index, such as the instanton number. In particular  this index is, up to an overall scale,  integer ${\cal P}_-\sim n \in{\mathbb Z}$ and the moduli space of BPS solutions  ${\cal M} $ is graded by $n$:
\begin{equation}
{\cal M}  = \oplus_{n\in \mathbb Z} {\cal M}_n  \ .
\end{equation}
Within each component ${\cal M}_n $ (apart from $n=0$) we see that $\{{\cal Q}_-,{\cal Q}_-\}\ne 0 $ and hence the ${\cal Q}_-$ supersymmetries are broken. Thus the resulting moduli space dynamics is only invariant under the eight ${\cal Q}_+$ supersymmetries. For $n=0$ the moduli space is flat and all sixteen supersymmetries are again realised.  Thus by embedding these one-dimensional sigma model dynamics in to a field theory we see that we are able to realise the full 16 supersymmetries and also make their higher-dimensional interpretation more transparent.

  \section{Conclusion}
  
  In this paper we have presented gauge theory actions in $2+1$ and $4+1$ dimensions (along with the dimensional reduction of the latter) without    boost invariance but with maximal supersymmetry. In particular some fields lack kinetic terms. As such one might be concerned that there is nothing to suppress them and the resulting theory will be pathelogical. However there are also Lagrange multiplier fields that restrict the dynamics to a moduli space of BPS configurations. As a result the kinetic energy of all the fields are controlled and the actions can be reduced to one-dimensional motion on the moduli space. This last step breaks half of the supersymmetry. One could state this result the other way around: we have managed to embed one-dimensional moduli space dynamics into a field theory and thereby double the supersymmetry and clarify the spacetime interpretation. 
 
These actions have been derived by solving the constraints of the $(2,0)$ system of \cite{Lambert:2010wm,Lambert:2016xbs} in the special null cases that were studied in \cite{Kucharski:2017jwv,Lambert:2011gb}. As such they are expected to describe limits of M2-branes and M5-branes where the branes have been infinitely boosted so that their worldvolume time coordinate becomes light-like. In other words in this construction these actions arise as a limit of of an infinite boost of static M2-branes and M5-branes, {\it aka} null M2-branes and M5-branes. Such embeddings  were discussed in \cite{Acharya:1998st} for the case of single branes. It is amusing to observe that the Lagrange multiplier fields $H$ and $G_{ij}$ which appear in our non-Lorentian actions both arise as components of the self-dual three-form of the six-dimensional $(2,0)$ supermultiplet. 
  
The sigma-models that result from the M5-brane action and its dimensional reduction are certainly not new. In particular for the uncompactified case they have appeared as a DLCQ prescription for the M5-brane $(2,0)$ SCFT \cite{Aharony:1997th,Aharony:1997an}. Indeed our result here provides another perspective on how this model relates to the $(2,0)$-Theory. We also expect that our action could be identified with  a non-abelian version of the M5-brane light-cone action  constructed in \cite{Bandos:2008fr}. In addition the AdS/pp-wave duals to these and similar DLCQ models was studied in \cite{Cvetic:1998jf,Brecher:2000pa,Maldacena:2008wh,Herzog:2008wg,Adams:2008wt} and it would be interesting to relate our construction in more detail to  these analyses.

It would also be interesting to derive these actions by taking  a non-Lorentzian scaling limit, perhaps something like a mixture of Carrollian and Galilean limits in the sense of 
\cite{Duval:2014uoa} (and \cite{Duval:2017els}  for pp-wave spacetimes), directly within the parent Lorentzian field theory without embedding the branes into eleven-dimensions. Or alternatively relate our modes  to the very special conformal symmetry models  constructed in
\cite{Nakayama:2018fib}.
Indeed one may expect that many supersymmetric field theories admit non-Lorentzian limits of this type which preserve all the supersymmetries and whose on-shell dynamics reduce to motion on a moduli space. Such a limit makes the Manton approximation where the dynamics are described by slow motion on a soliton moduli space exact. It also raises the question of what is the classification of all field theories with 16 supersymmetries if one does not impose the condition of Lorentz invariance.

\section*{Acknowledgements}
We would like to thank S. Muhki for helpful discussions. 
  N. Lambert is supported in part by STFC grant
 ST/P000258/1 and M. Owen is supported by the STFC studentship ST/N504361/1.
 
  \section*{Appendix: Fierz Identities}
  
Here we list some identities that arise from the Fierz identity. In section two one has the following:
\begin{align}
0&=\langle\Psi^T_+,[X^I,(\epsilon^T_-\hat\Gamma_{Z{\bar Z}}\hat\Gamma^I\Psi_+), \Psi_+]\rangle +\langle\Psi^T_+,[X^I,(\epsilon^T_-\hat\Gamma^J\Psi_+), \hat\Gamma_{Z{\bar Z}}\hat\Gamma^{IJ}\Psi_+]\rangle \nonumber\\
0&=\langle\Psi^T_+,[X^I,(\epsilon^T_+\hat\Gamma^I\hat\Gamma_{\bar z}\Psi_+),\hat\Gamma_{z}\Psi_-]\rangle -\langle\Psi^T_+,[X^I,(\epsilon^T_+\hat\Gamma^I\hat\Gamma_z\Psi_+),\hat\Gamma_{{\bar z}}\Psi_-]\rangle \nonumber\\ 
&\quad+\langle\Psi^T_+,[X^I,(\epsilon^T_+\hat\Gamma^I\hat\Gamma_{Z}\Psi_+) ,\hat\Gamma^I\hat\Gamma_{\bar Z}\Psi_-]\rangle- \langle\Psi^T_+,[X^I,(\epsilon^T_+\hat\Gamma^I\hat\Gamma_{\bar Z}\Psi_+), \hat\Gamma^I\hat\Gamma_{Z}\Psi_-]\rangle\nonumber\\ 
0&=\langle\Psi^T_+,[Z,(\epsilon^T_-\hat\Gamma_Z\Psi_-), \Psi_+]\rangle +2\langle\Psi^T_+,[Z,(\epsilon^T_-\hat\Gamma_{{\bar z} Z}\Psi_+), \hat\Gamma_z\Psi_-]\rangle\nonumber\\
&\quad -\langle\Psi^T_+,[Z,(\epsilon^T_-\hat\Gamma^I\Psi_+),\hat\Gamma^I \hat\Gamma_Z\Psi_-]\rangle \nonumber\\ 
0&=\langle\Psi^T_-,[Z,(\epsilon^T_+\hat\Gamma_Z\Psi_+), \Psi_-]\rangle +2\langle\Psi^T_-,[Z,(\epsilon^T_+\hat\Gamma_{{  z} Z}\Psi_-), \hat \Gamma_{\bar z}\Psi_+]\rangle\nonumber\\
&\quad -\langle\Psi^T_+,[Z,(\epsilon^T_+\hat\Gamma^I\Psi_-),\hat\Gamma^I \hat\Gamma_Z\Psi_+]\rangle \ .
\end{align}
There are also similar identities where $Z\to \bar Z$.
These can be derived from the vanishing of the cubic fermion terms that arise in $\delta S$ for the maximally supersymmetric M2-brane theory  and then splitting-up the fields into their various components, {\it e.g.} $\Psi=\Psi_++\Psi_-$, $\epsilon = \epsilon_++\epsilon_-$, $X^I\to X^I,Z,\bar Z$,  where the sign indicates their chirality with respect to $\hat\Gamma_{034}$.

In section three the following Fierz identities  arise:
\begin{align}
0&={\rm tr}\left(\Psi^T_-[(\epsilon^T_- \Gamma_0\Psi_+), \Psi_-]\right)+{\rm tr}\left(\Psi^T_+[(\epsilon^T_-  \Gamma_m\Psi_-),\Gamma_0\Gamma_m \Psi_+]\right)\nonumber\\
&+{\rm tr}\left(\Psi^T_-[(\epsilon^T_-  \Gamma_m\Psi_-),\Gamma_0\Gamma_m \Psi_+]\right)
+{\rm tr}\left(\Psi^T_-[(\epsilon^T_-  \Gamma_0\Gamma_m\Psi_+), \Gamma_m \Psi_-]\right)
\nonumber\\
0&={\rm tr}\left(\Psi^T_-[(\epsilon^T_+ \Gamma_0\Psi_-), \Psi_-]\right)-{\rm tr}\left(\Psi^T_-[(\epsilon^T_+ \Gamma_0 \Gamma_m\Psi_-),\Gamma_m \Psi_-]\right)
\ ,
\end{align}
where $m=1,2,3,4,6,7...,10$ ({\it i.e.} $m\ne 5$).
These can be derived from the vanishing of the cubic fermion terms that arise in $\delta S$ in   five-dimensional  maximally supersymmetric Yang-Mills theory and then  splitting-up the fields into their various components, {\it e.g.} $\Psi=\Psi_++\Psi_-$, $\epsilon = \epsilon_++\epsilon_-$ where the sign indicates their chirality with respect to $\Gamma_{05}$.

\end{document}